\documentclass{article}
\usepackage{makeidx,epsfig}
\usepackage{setspace,graphicx}%,subfigure

\usepackage{amsfonts}
\usepackage{amsmath}
\usepackage{hyperref}
\usepackage{Generic}

\usepackage{tikz}
\usepackage{subcaption}
\usepackage{caption}
\usetikzlibrary{positioning,shapes.misc}

\begin{document}

\begin{doublespace}

~
\vskip 2cm

\noindent {\huge \bf Hockey Player Performance via

\noindent Regularized Logistic Regression} 

\vskip .5cm
\noindent
{\it \Large Robert B.~Gramacy, Matt Taddy, and Sen Tian}

\vskip 2cm
A hockey player's plus-minus measures the difference between goals
scored by and against that player's team while the player was on the ice.
This measures only a marginal effect, failing to account for the influence of
the others he is playing with and against.   A better approach would be to
jointly model the effects of all players, and any other confounding
information, in order to infer a partial effect for this individual: his
influence on the box score regardless of who else is on the ice.

This chapter describes and illustrates a  simple algorithm for recovering such
partial effects.  There are two main ingredients.  First, we provide a
logistic regression model that can predict which team has scored a given goal
as a function of who was on the ice, what teams were playing, and details of
the game situation (e.g. full-strength or power-play).  Since the resulting
model is so high dimensional that standard maximum likelihood estimation
techniques fail,  our second ingredient is a scheme for regularized
estimation.  This adds a penalty to the objective that favors parsimonious
models and stabilizes estimation.  Such techniques  have proven useful in
fields from genetics to finance over the past two decades, and have
demonstrated an impressive ability to gracefully handle large and highly
imbalanced data sets.  The latest software packages accompanying this new
methodology -- which exploit parallel computing environments, sparse matrices,
and other features of modern data structures -- are widely available and make
it straightforward for interested analysts to explore their own models of
player contribution.

This framework allows us to quickly obtain high-quality estimates for the full
system of competing player contributions. After introducing the  measurement
problem in Section \ref{sec:intro}, we detail our regression model in Section
\ref{sec:regression} and the regularization scheme in Section
\ref{sec:regularization}.  The remainder of the chapter analyzes more than a
decade of data from the NHL.  We fit and interpret our main model, based on
prediction of goal scoring, in Section \ref{sec:goals}.  This is compared to
shot-based analysis, and metrics analogous to Corsi or Fenwick scores, in
Section \ref{sec:shots}.  Finally, Section \ref{sec:salary} considers the
relationship between our estimated  performance scores and  player salaries.
Overall, we are able to estimate a partial plus-minus metric that occurs on
the same scale as plus-minus but controls for other players and confounding
variables.  This metric is shown to be more highly correlated with salary than
the standard (marginal) plus minus.  Moreover, we find that the goals-based
metric is more correlated with salary than those based upon shots and blocked
shots.   We conclude in Section \ref{sec:conclusion} with thoughts on further
extensions, in particular by breaking out of the linear framework to use
classification models popular in the Machine Learning literature.

The code for all empirical work in this chapter is provided to the public via a GitHub
repository (\verb!https://github.com/TaddyLab/hockey!) and utilizes open
source libraries for {\sf R} \cite{cranR}, particularly the {\sf gamlr} \cite{gamlr} package from \cite{taddy_one-step_2015}.

\section{Introduction: marginal and partial effects}
\label{sec:intro}

Hockey is played on ice, but that's not all that sets it apart from seemingly
related  sports like soccer, basketball, or even field hockey.  At least not
from an analytics perspective.  The unique thing about hockey is the rapid
substitutions transpiring continuously during play, as well as at stoppages in
play.  In the data sets we have compiled, which we discuss in more detail
shortly, the median amount of time observed for a particular on-ice player
configuration (determined by unique players on the ice for both teams) is a
mere eight seconds. Although many ``shifts'' are much longer than that, a
trickle of piecemeal substitutions on both sides, transpiring as play
develops, makes it difficult to attribute credit or blame to players for
significant events, such as goals or shots.

Plus-minus (PM) is a traditional metric for evaluating player contributions in
hockey. It is calculated as the difference, for a given player, between the
number of goals scored against the player's team and those scored by the
player's team while that player was on the ice.  For example, during the
2012-2013 season Stanley Cup Finals, between Boston and Chicago, Duncan Keith
of the Chicago Blackhawks was on the ice for 8 goals by Chicago and 4 by
Boston, giving him a +4 PM for the series.
% c(-1,1)[Y[64538:64569]+1]*XP[64538:64569,"DUNCAN_KEITH"]

The PM score represents what statisticians call a \textit{marginal effect:}
the average change in some response (goals for-vs-against) with change in some
covariate (a player being on the ice) \textit{without accounting for whatever
else changes at the same time}. It is an aggregate measure that averages over
the contributions of other factors, such as teammates and opponents.   For
example, suppose that the three authors of this chapter are added to the
Blackhawks roster and that Joel Quenville (the coach of the Blackhawks) makes
sure that Duncan Keith is with us on the ice whenever we are playing.  Since
none of us are anything close to as good at hockey as Keith is, and surely our
poor play would allow the other team to score, this will cause Duncan Keith's
PM to drop.  At the same time, our PMs will be much higher than they would be
if we didn't get to play next to Duncan Keith.

Due to its simplicity and minimal data requirements, plus-minus has been a
preferred metric for the last fifty-odd years.  But since it measures a
marginal effect, the plus-minus is impacted by many factors beyond player
ability, which is the actual quantity of interest.  The ability of a player's
teammates, or the quality of opponents, are not taken into account.  The amount of
playing time is also not factored in, meaning plus-minus stats are much
noisier for some players than others. Finally, goalies, teams, coaches,
salaries, situations, special teams, etc.~-- which all clearly contribute to
the nature of play, and thus to goals -- are neither accounted for when
determining player ability by plus-minus, and nor are they used to explain
variation in the goals scored against a particular player or team.

Instead of marginal effects, statisticians are more often interested in
\textit{partial effects}: change in the expected response that can be
accounted for by change in your variable of interest \textit{after removing
the change due to other influential variables.} In the example above, a
partial effect for Duncan Keith would be unchanged if he plays with the
authors of this article or with the current members of the Blackhawks.  In
each case, the partial effect will attempt to measure how Duncan Keith can
influence the box-score regardless of with whom he skates. Because such partial
effects help us predict how Keith would perform on a different team or with a
different combination of line-mates, this information is more useful than
knowing a marginal effect.

One way that statisticians can isolate partial effects is by running
\textit{experiments}.  Suppose that now, instead of playing for the
Blackhawks, we are coaching them.  In order to figure out the value of Keith,
we could randomly select different players to join him whenever he is on the
ice and send completely random sets of players onto the ice whenever he is not
playing.  Then, due to the setup of this randomized experiment, Keith's
resulting PM score will represent a partial effect -- his influence regardless
of who he plays with.  Of course, no real hockey coach would ever manage their
team in this way.  Instead, we hockey analysts must make sense of
\textit{observational data} that is collected as the games are naturally
played, with consistent line mates and offensive-defensive pairings and where
Duncan Keith tends to play both with and against the best players available.

Partial effects are measured from observational data through
\textit{regression}: you model the response (e.g., goals) as a function of
many  influential variables (\textit{covariates}; e.g., all of the players on
the ice).  With rich enough data, we can simultaneously estimate the full set
of competing partial effects corresponding to all of our influential
variables.  This is straightforward when there are only a small number of
covariates.  However, the standard regression algorithms will fail when the
number of covariates is large. This `high dimensional regression' setting
occurs in hockey analysis, where we would like to regress `goals' onto the set
of variables corresponding to whether each NHL player is on the ice (a set of
2500 players in our dataset) while also including effects of team, season,
playoffs, and special teams scenarios (e.g. power plays).  Moreover, the
covariate design is highly \textit{imbalanced}: over the span of several
seasons there may be  tens of thousands of goals, but players play with and
against only a small fraction of other players and the number of unique player
configurations is relatively small.  Due to the use of player lines, and
consistent line match-ups with opponents, where groups of two or three players
are consistently on ice together at the same time, the data contain many
clusters of individuals who are seldom observed apart.  Standard regression
algorithms, such as maximum likelihood inference via Fisher scoring, will
either massively \textit {over-fit} (e.g. assign large effects to players who
rarely play) or simply fail to converge.

However, there has been a tremendous improvement over the past two decades in
the techniques available for high dimensional regression analysis.  These
advancements are driven by the demands of researchers in genetics and finance,
for example, for whom resolving partial effects amongst  large sets of
variables is the key to their science.  The most successful approaches
introduce some amount of \textit{regularization}  to the estimation
problem -- an additional penalty term that rewards simplicity (e.g.,
\cite{hastie:tibsh:fried:2001}).  In our context, regularization shrinks
towards a model where individual players don't make a huge difference while
still allowing for large estimated player effects when the data warrant it.
This conforms to what most analysts already believe: many players have a
neutral, or ``zero'', effect (relative to the NHL average), whereas some are
stars and others are liabilities.     The amount of regularization is chosen
to make the model perform as well as possible in out-of-sample prediction and,
again, contemporary statistical learning tools are designed to do exactly
this -- reliably predict the future. To take advantage of these tools, we  need
only to phrase partial player effect estimation  as a regression problem.

\section{Regression Model}
\label{sec:regression}

The goal of our regression analysis will be to estimate a model that  relates
individual presence on the ice to observable outcomes of interest.  We
describe the model here for a goals-based analysis, but extend it to shots and
other metrics in the analysis sections.

Previous attempts at partial player effect estimation range from standard
linear regression (usually on aggregate data) -- the adjusted plus-minus
scores of \cite{awa09}, \cite{schlocwel10},  and \cite{mac10} -- to the complex
hazard model of
\cite{ThoVenJen12}, which proposes a proportional hazards process for
game events, allowing partial player effects to be backed out from high
resolution game data. 
Adjusted plus-minus is built from similar ideas for Basketball analysis (see
\cite{ros04} and \cite{ilabar08}). Its linear model analysis implies an
underlying normality assumption for the error structure; this may be a good
approximation for basketball, where scoring is frequent and variability in
player configurations is small, but it is  inappropriate for disaggregated
data with a binary response (e.g., whether an individual goal is
for-vs-against the home team).  Such misspecification becomes especially
problematic when combined with the modern regularization techniques necessary
for reliable estimation of high dimensional models.  On the other hand, more
complex stochastic process modeling requires many additional assumptions on
the data generating process and can be difficult to validate in practice;
moreover, models such as that of \cite{ThoVenJen12} take far longer to run
than we wish for our analysis.  Some other important contributions to
estimating player ability and attributing that to team success include
\cite{stair:etal:2011,pettigrew:2015,mason:foster:2007}.

The goal of our modeling is to provide a correct treatment of the binary
`goals' data without introducing significant additional modeling complexity.
In particular, we advocate the simple \textit{logistic regression} framework
suggested by \cite{gramacy:jensen:taddy:2013}.  In logistic regression, the
average log odds of a goal being scored ``for'' a particular team  is modeled
as a linear function of predictor variables which may be comprised of an
indicator of player configuration and other quantities, which is otherwise
identical to the familiar ordinary (least squares) regression setup.  We
provide a detailed description of the model here, but refer the reader to
\cite{sheather:2009} or similar texts covering {\em Generalized Linear Models
(GLMs)}, of which logistic regression is a special case.
The setup is rather straightforward, easy to
extend, and highly interpretable.  Estimated coefficients describe
contributions to the log odds of goals, and we show that these can be
converted back onto the scale of goals, resulting in a adjusted plus-minus
statistic, but this time one which is a true partial effect.  

Given $n$ goals throughout the National Hockey League (NHL) over some
specified time period,  say $y_i$ is $+1$ for a goal by the \textit{home} team
and $-1$ for a goal by the \textit{away} team.\footnote{{\em home} and {\em
away} are merely organizational devices, creating a consistent binary
bifurcation for goals that can be applied across games, seasons, etc. Due to
the symmetry in the logit transformation, player effects are unchanged when
framing away team probabilities as $q_i$ rather than $1-q_i$, so we loose no
generality by privileging home team goals in this way.}  Say that $\boldsymbol{q_i =
\mathrm{p}(y_i = 1) =  \mathrm{p}(\textbf{home~team~scored~goal}~i)}$.  The
logistic regression model of player contribution is, for goal $i$ in season
$s$ with away team $a$ and home team $h$,  
\begin{align}\label{hockeymod}
\log\left[\frac{q_{i}}{1-q_{i}}\right] = \alpha + \mathbf{u}_i'\boldsymbol{\gamma} +
\mathbf{v}_i'\boldsymbol{\varphi} + \mathbf{x}_i'\boldsymbol{\beta}_0 +
(\mathbf{x}_i\circ\mathbf{s}_i)'(\boldsymbol{\beta}_s + p_i \boldsymbol{\beta}_{p}), 
\end{align}  
where
\begin{itemize}
\item Vector $\mathbf{u}_i$ holds indicators for each team-season (e.g., the Blackhawks in 2012-2013 would correspond to a coordinate of $\mathbf{u}_i$), set 
$u_{it}=+1$ if team-season $t$ was the home team for goal $i$, $u_{it}=-1$ for the away team, and $u_{it}=0$ if team-season $t$ was not on the ice for goal $i$.  This information is included to control for factors beyond the player's control, such as quality of coaching and fan support.
\item Vector $\mathbf{v}_i$ holds indicators for various special-teams scenarios (e.g., being short-handed on a penalty kill), again set $v_{ik}=+1$ if the home team is in special-teams scenario $k$ when goal $i$ was scored, $v_{ik}=-1$ if the away team is in scenario $k$, and $v_{ik}=0$ if neither team was in scenario $k$ when goal $i$ was scored.  We consider $6$ non-six-on-six settings (6v5, 6v4, 6v3, 5v4, 5v3, 4v3) and an additional `pulled goalie' indicator; note that more than $35\%$ of the goals occur on some type of special teams scenario. 
\item Vector
$\mathbf{x}_i$ contains player-presence indicators, set $x_{ij}=1$ if player $j$ was on the
home team and on ice for goal $i$, $x_{ij}=-1$ for away player $j$ on ice for
goal $i$, and $x_{ij}=0$ for everyone not on the ice. With $\circ$ denoting the Hadamard (element-wise) product, this player vector is also interacted with 
\begin{itemize}
\item  season (e.g., 2013-2013) vector $\mathbf{s}_i$, with $s_{ti} = 1$ if goal $i$ was scored in season $t$, and
\item the post-season indicator $p_i$  for whether or not the goal was scored in the playoffs, with $p_{i} =1$ for the playoffs and zero for the regular season.
\end{itemize}  
By interacting players with seasons and with playoffs in this way, we have the
potential to differentiate player ability over time both within (regular
v.~post-) season(s) and across seasons.  There is potential for confounding
with team--season effects $\mathbf{u}_i$ however, as very few players change
teams during a season.

\end{itemize}
In this full specification the number of parameters, i.e., $K =
|\boldsymbol{\alpha}|+ | \boldsymbol{\phi}| + |\boldsymbol{\beta}_0| +
|\boldsymbol{\beta}_s| + |\boldsymbol{\beta}_p|$ is on the order of the number
of players, $p$, in the league spanning the seasons/games of interest.  The
exact number depends on modeling aspects, like the number of special teams
scenarios (i.e., constant in $p$), and on quantities like the number of
team-seasons which grow more slowly than $p$.

To explain the coefficients and their interpretation, 
$\beta_{0j} + \beta_{sj}$ is the regular-season-$s$ effect of player $j$ on the log
odds that, given a goal has been scored, the goal was scored by their team.  Coefficient 
$\beta_{0j} + \beta_{sj} + \beta_{pj}$ is the corresponding effect for post-season-$s$ (note that, under the regularization scheme in the next section, $\beta_{pj}$ will be fixed at zero unless player $j$ reaches the playoffs). 
These effects are `partial' in that they control for who else was on the ice,
special teams scenarios, and team-season effects -- a player's
$\beta_{0j}$ or $\beta_{sj}$ only need be nonzero if that player effects play
above or below the team average for a given season.  A test of
understanding: what does the intercept $\alpha$ represent in
(\ref{eq:simplemodel})?\footnote{It is the home ice advantage: if you do not
know anything about who is playing or on the ice, the odds are $e^{\alpha}$
higher that the home team has scored any goal.}

For intuition, consider a simple ``player-only'' version of our model that has
only who-is-on-the-ice as a time-invariant influence on goal scoring.  This is the version of the model that was applied in \cite{gramacy:jensen:taddy:2013}.  Then
there are no team-season-specific intercepts ($\alpha_{sh}=\alpha_{sa}=0$), no
special teams effects ($\boldsymbol{\phi}=\mathbf{0})$, and no season-specific
player-effect changes ($\boldsymbol{\beta}_s = \mathbf{0}$ and $\boldsymbol{\beta}_p = \mathbf{0}$)  so that $\beta_j
= \beta_{0j}$ is the constant effect of player $j$.  The log odds that the
home team has scored a given goal become \begin{equation} \log
\left[\frac{q_i}{1-q_i} \right]  = \alpha + \beta_{h_{i_1}} + \cdots +
\beta_{h_{i_6}} -  \beta_{a_{i_1}} - \cdots - \beta_{a_{i_6}},
\label{eq:simplemodel} \end{equation} where the subscripts on the coefficients
$\beta$ are as follows: $h_{i_1}, \dots, h_{i_6}$ are the six players on the
ice for the home team and  $a_{i_1}, \dots, a_{i_6}$ indicate the players for
the away time.\footnote{In this setup the goalies {\em are} included in the
calculations, unlike with {\em plus-minus}.}  This is the just a re-writing of
$\mathbf{x}_i'\boldsymbol{\beta}$ from (\ref{hockeymod}), where the vector
$x_i$ (of length equal to the number of players) contains the ``$+1$'' and
``$-1$'' indicators depending on whether that player was on the home or away
team, and where all other $x_{ij}$ are zero so that $\sum_j |x_{ij}| = 12$ for
full-strength play. See Figure \ref{fig:data} for illustration. 

\begin{figure}[hbt]
    \centering
    \begin{tikzpicture}
    \matrix [column sep=7mm, row sep=2mm] {
        \node [draw=none, fill=none] {$Y$: scoring team}; &
        \node [draw=none, fill=none] {$X_P$: players}; &\\
        \node [draw=none, fill=none] {$y_i\in\{{-1},1\}$}; &
        \node [draw=none, fill=none] {$x_{Pij}\in\{{-1},0,1\}$}; &\\
        \node (y1) [draw, shape=rounded rectangle] {$\hspace{7pt}1$}; &
        \node (xp1) [draw, shape=rounded rectangle, label={[label distance=0.1mm]90:\small $1 \hspace{190pt} n_p$}] {$0\, 1\, {-1}\, 1\, 0\, {-1}\, 0\, {-1}\, 1\, {-1}\, {-1}\,  0\, 0 \cdots 0\, 1\, 0\, 1\, 1$}; &\\
        \node [draw=none, fill=none] {$\vdots$}; &
        \node [draw=none, fill=none] {$\vdots$}; &\\
        \node [draw, shape=rounded rectangle] {$-1$}; &
        \node [draw, shape=rounded rectangle] {$1\, {-1}\, 1\, {-1}\, 0\, 1\, 0\, 1\, {-1}\, 1\, 1\, 0 \cdots 0\, {-1}\, 0\, {-1}\, {-1}$};&\\
    };
    \end{tikzpicture}
    \caption{Diagram of a simple design matrix for a `players only model' and two example goals (rows). The two goals are shown in the same season under the same configuration of teams except that the first goal was scored by the home team while the second goal was by the away team. The configurations of players are only differed by the first player since the home team was on a 6v5 power play for the first goal. }\label{fig:data}
\end{figure}

The model in \ref{hockeymod} is simple and transparent; if you wish to
control for new variables or situations you just need to add covariates to the
logistic regression. In theory, one could fit the model easily in  {\sf R} by
typing \begin{verbatim} R> fit <- glm(y~X, family="binomial") \end{verbatim}
But unfortunately, that just doesn't work.   The problem is that hockey data
is too high dimensional (too many covariates) for ordinary logistic regression
software, and the design matrices are too highly imbalanced to obtain
meaningful (low variance) estimates of player effects.   In almost all
regressions one is susceptible to the temptations of rich modeling when the
data set is large, and our hockey setup is no different. One must be careful
not to {\em over-fit}, wherein parameters are optimized to statistical noise
rather than fit to the relationship of interest.  And one must be aware of
{\em multicollinearity}, where groups of covariates are correlated with each
other making it difficult to identify individual effects, as happens when
players are grouped into lines.

A first approach to finding a remedy might be to entertain stepwise
regression, e.g., via {\tt step} in {\sf R} with a stopping rule based upon an
information criteria like AIC and BIC.  But that also doesn't work on this
data: the calculations take days, and turn up very few non-zero predictors
(i.e., players whose presence have any effect on goals).  The trouble here is
that players  can't be judged on their own, since they almost always play with
and against eleven others.  Therefore the one-at-a-time judgments made by {\tt
step} fail to discover many relevant players despite making a combinatorially
huge number of such comparisons.  Moreover, stepwise regression results are
well known to be highly variable: tiny jitter to the data can lead to massive
changes in the estimated model. The combined effect is an unstable algorithm
that yields overly simple results and takes a very long time to run.

Instead, a crucial contribution of
\cite{gramacy:jensen:taddy:2013} is to suggest the use of modern penalized and
Bayesian logistic regression models, which biases the estimates of player
effects towards zero. In the next section we consider one fast and successful
version of these methods: $L_1$ regularization.

\section{Regularized Estimation}
\label{sec:regularization}

\begin{figure}[t]
\includegraphics[width=\textwidth]{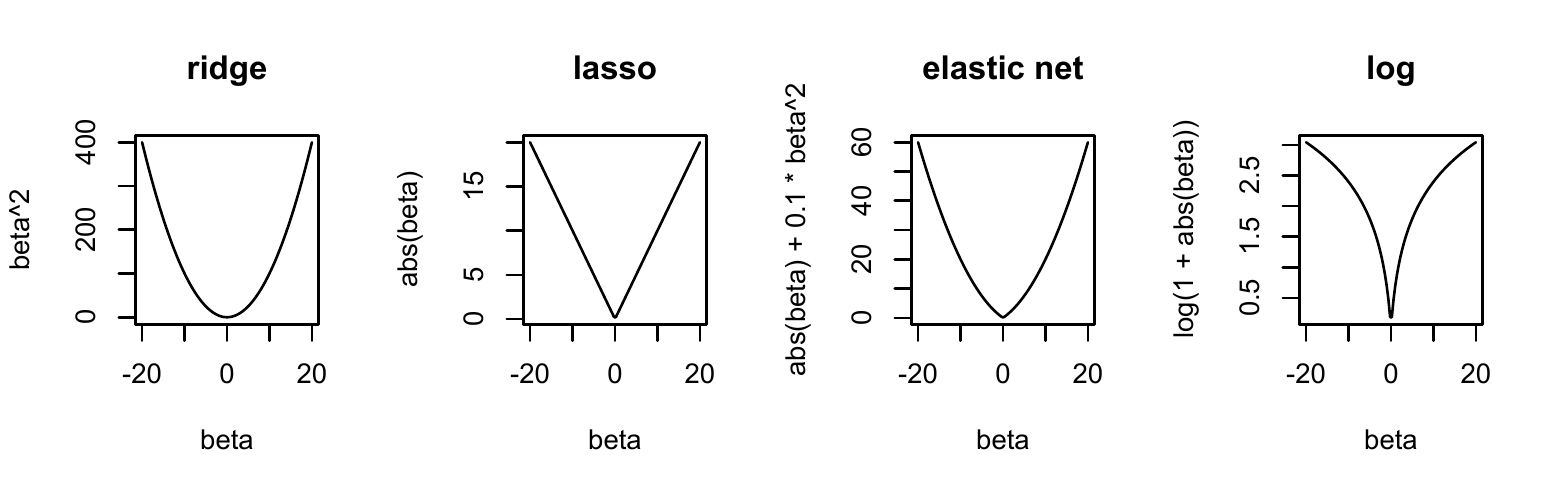}
\caption{\label{costs} 
From left to right, 
$L_2$ `ridge' costs \cite{hoerl_ridge_1970}, $L_1$ `lasso' \cite{tibshirani_regression_1996}, the `elastic net' mixture of $L_1$ and $L_2$ \cite{zou_regularization_2005}, and the log penalty \cite{candes_enhancing_2008}.
}
\end{figure}

Our solution is to take a modern regularized approach to regression.  If
$\eta_i = \log[ q_i/(1-q_i) ]$ is our linear equation for log odds from
(\ref{hockeymod}), then the usual maximum likelihood estimation routine (e.g.,
via {\sf glm} in {\sf R}) minimizes the negative log likelihood objective
for $n$ goal events
\begin{equation}
\label{eq:nllhd} l\left(\boldsymbol{\eta}; \mathbf{y}\right) =
\sum_{i=1}^n \log\left(1 + \exp[-y_i \eta_i]\right). 
\end{equation} 
Instead, a
\textit{regularized} regression algorithm will minimize a penalized objective,
say for example $l\left(\boldsymbol{\eta}; \mathbf{y}\right) + n\lambda
\sum_{j=1}^p\left[ c_j\left(|\beta_{0j}|\right) +
c\left(|\beta_{sj}|\right)\right]$, where $\lambda>0$ controls overall penalty
magnitude, $c_j(\cdot)$ are coefficient cost functions, while $n$ is the total
number of goals and $p$ is the number of players.
A few common cost functions are shown in Figure \ref{costs}.  Those that have
a non-differentiable spike at zero (all but ridge) lead to sparse estimators,
meaning that many coefficients are set to exactly zero.   The curvature of the penalty
away from zero dictates the weight of shrinkage imposed on the nonzero
coefficients:  $L_2$ costs increase with coefficient size,  lasso's $L_1$
penalty has zero curvature and imposes constant shrinkage, and as curvature
goes towards $-\infty$ one approaches the $L_0$ penalty of subset selection.

In this article we focus on the $L_1$ penalty for its balance between
shrinkage of large signals (players tend not to have huge effects) and a
preference for sparsity (we can only measure the nonzero effects of a subset
of players).  For these and many appealing theoretical reasons (and for
computational tractability), the $L_1$ penalty is by far the most commonly
used in contemporary regularized regression; see
\cite{hastie:tibsh:fried:2001} for a broad-audience overview and
\cite{taddy_one-step_2015} for details on the algorithms used in this chapter.
Under lasso $L_1$ penalization, estimation for the unknown parameters in 
our particular hockey model (\ref{hockeymod}) proceeds through optimization of
\begin{equation} \label{pendev}
l\left(\boldsymbol{\eta}; \mathbf{y}\right) + n\lambda \sum_{j=1}^p\left(|\beta_{0j}| + |\beta_{sj}| + |\beta_{pj}|\right).
\end{equation}
It is important to note there that we are \textit{penalizing only the player
effects}.  The team-season effects ($\boldsymbol{\gamma}$) are unpenalized.
This strategy of combining penalized and unpenalized estimation is advocated
in, e.g.,
\cite{taddy_distributed_2015} and \cite{gentzkow_measuring_2015}.  It works
nicely whenever you have a subset of covariates for which there is strong data
signal (many repeated observations, which we have for team-season and special
teams effects) and whose effect you'd like to completely remove from
estimation for other coefficients.  In this way, we ensure that the player
effect estimates are not \textit{polluted} by confounding effects in
$\mathbf{u}$ and $\mathbf{v}$.  

Moreover, consider the covariates on $\boldsymbol{\beta}_0$,
$\boldsymbol{\beta}_s$, and $\boldsymbol{\beta}_p$ in (\ref{hockeymod}): for the latter two,
 $\mathbf{x}_i$ interacts with additional binary indicators, such that
$\beta_{0j}$ acts on more nonzero terms than $\beta_{sj}$, which itself acts
on more nonzero terms than $\beta_{pj}$.  Thus there is  less signal
associated with the season and playoff player effect innovations than with the
player baseline effects, so that these will tend to be estimated at zero unless
there is significant evidence that a player has become better or worse across
seasons or in a given post-season.

Penalty size, $\lambda$, acts as a {\it squelch}: canceling noise to
focus on the true input signal. Large $\lambda$ lead to very simple  model
estimates, while as $\lambda \rightarrow 0$ we approach maximum likelihood
estimation. Since you don't know optimal $\lambda$, practical
application of penalized estimation requires a {\it regularization path}: a $p
\times T$ field of $\boldsymbol{\hat\beta}$ estimates obtained while moving
from high to low penalization along $\lambda^1 > \lambda^2 \ldots >
\lambda^T$. These paths begin at $\lambda^1$ set to infimum $\lambda$ such
that (\ref{pendev}) is minimized at $\boldsymbol{\hat\beta} = \mathbf{0}$, and proceed
down to some pre-set $\lambda^T$ (e.g., $\lambda^T= 0.01\lambda^1$).

 A common tool for choosing the optimal $\lambda$ -- that for which we report
 estimated player effects -- is cross validation (CV).  In CV, the path of
 coefficients is repeatedly fit to data subsamples and used to predict the
 response on the left-out data. The $\lambda$ leading to minimum error is then
 selected as optimal.  In this chapter, we instead use an analytic alternative
 to CV that yields models that perform as well or better out-of-sample.   The
 corrected Akaike Information Criterion (AICc), proposed in
 \cite{hurvich_regression_1989}, is defined as
\begin{equation*}
AICc = 2\sum_{i=1}^{n} l(\boldsymbol{\hat\eta}_{\lambda};\mathbf{y}) + \frac{2kn}{n-k-1},
\end{equation*}
where $\boldsymbol{\hat\eta}_{\lambda}$ are the estimated log odds under
penalty $\lambda$ and $k \leq K$ is the number of non-zero estimated
coefficients (likely far fewer than the total number of parameters $K$, as
many players cannot be distinguished from the league average) at this penalty.
See \cite{taddy_one-step_2015} and \cite{flynn_efficiency_2013} for details on
AICc selection in this context; we find AICc preferable to CV because it is
computationally efficient (you only need to optimize once) and because there
is no random Monte Carlo variation -- it always gives the same answer on the
same data.  However, all of our ideas here apply if you wish to use CV
selection instead.

The {\tt gamlr} package \cite{gamlr} can
be called via {\sf R} to implement this procedure:
\begin{verbatim}
R> fit <- gamlr(X, Y, standardize=FALSE, family="binomial")
\end{verbatim}
For CV, just replace \verb!gamlr! with \verb!cv.gamlr!. The
\verb!standardize=FALSE! flag tells \verb!gamlr! to \textit{not} weight the
coefficient penalties by the standard deviation of the corresponding covariate
(i.e., to use penalty $\lambda|\beta_j|$ instead of
$\lambda\,\mathrm{sd}(\mathbf{x}_j)|\beta_j|$); this is appropriate here because
such standardization would \textit{up-weight} the influence of players who
rarely play (and have low $\mathrm{sd}(\mathbf{x}_j)$) relative to those who
have a lot of ice time (and thus high $\mathrm{sd}(\mathbf{x}_j)$).  The
software exploits sparsity in our player effects (\verb!X!) via the
\verb!Matrix! library for {\sf R}, and is extremely fast to run: no examples
in this article require more than a few seconds of computation. Estimated
coefficients at optimal $\lambda$ are available as \verb!coef(fit)!.

One natural way to understand regularized regression is through the lens of
Bayesian posterior inference. Judiciously chosen prior distributions lend
stability to the fitted model, which is crucial in contexts where the number
of quantities being estimated is large. In our setting where larger
$\beta$-values indicate large positive or negative contributions to player
ability, it makes sense to choose a prior that encourages coefficients to
center around zero, a so-called {\em shrinkage} prior.  Our {\em a priori}
belief is that most players are members of the rank-and-file: their
contribution to goals is {\em neutral} (e.g., zero on the log-odds scale), and
that only a handful of stars (and liabilities) have a strong contribution to
the chances of scoring (or letting in) goals.  From the perspective of point
estimation, adding a prior on $\beta_j$ centered at zero is equivalent to
adding a penalty term for $\beta_k \ne 0$ in our objective function. Different
choices of priors correspond to different penalty functions on $\beta_j \ne
0$; a Laplace prior distribution on each $\beta_j$  corresponds to our $L_1$
penalty in (\ref{pendev}). The posterior density is the product of likelihood
and prior, and on the log scale that product becomes a a sum. So maximizing
the posterior to obtain posterior modes is equivalent to maximizing the log
likelihood plus a penalty term which is the log of the prior. Conversely, 
minimizing (\ref{pendev}) may be interpreted as Bayesian
posterior maximization.

\section{Analysis: goal-based effects}
\label{sec:goals}

This section attempts to quantify the performance of hockey players using data
from the NHL. It extends the analysis in \cite{gramacy:jensen:taddy:2013}, which used
a smaller dataset, assumed constant player effects, and did not control for
team-season effects. The data, downloaded from \url{http://www.nhl.com},
comprise of play-by-play NHL game data for regular and playoff games during
$11$ seasons of 2002-2003 through 2013-2014\footnote{Season 2004-2005 was a
lockout that resulted in a cancellation}. The data capture all signifigant
events in every single game, such as change, goal, shot, blocked shot, miss
shot, penalty and etc. There were $p=2439$ players involved in $n=69449$
goals.

The analysis proceeds through estimation of the model from (\ref{hockeymod}),
\begin{align*}
\log\left[\frac{q_{i}}{1-q_{i}}\right] = \alpha + \mathbf{u}_i'\boldsymbol{\gamma} +
\mathbf{v}_i'\boldsymbol{\varphi} + \mathbf{x}_i'\boldsymbol{\beta}_0 +
(\mathbf{x}_i\circ\mathbf{s}_i)'(\boldsymbol{\beta}_s +
p_i'\boldsymbol{\beta}_{p}), \end{align*} by minimization of the implied
penalized deviance in (\ref{pendev}). The estimated player coefficients for
each season $s$ are then available as $\beta_{0j} + \beta_{sj}$,  the
combination of a  baseline effect plus a season-specific innovation.  These
effects represent the estimated change in the log odds that, given a goal was
scored, the goal was scored by player $j$'s team.

The estimated effects -- our $\beta_{0j}$ and $\beta_{sj}$ -- might be tough for
non-statisticians to interpret. One option is to translate from the scale of
log-odds to that of probabilities. In particular, we define the `partial for-\%' functional as
\begin{equation}\label{eq:pfp}
\text{PFP}_{sj} = \left(1 + \exp[-\beta_{0j} - \beta_{sj}]\right)^{-1}.
\end{equation}
This feeds the player effects through a logit link  to obtain  the
probability that, given a goal was scored in season $s$, it was scored by
player $j$'s team, {\it if we know nothing else other than that player $j$ was
on the ice}. It lives on the same scale as the commonly used \textit{for-\%}
(FP) statistic: the total number of events by a given player's team
\textit{divided} by the total number of events by either team, while that
player was on the ice.   Like PM, FP is a marginal effect that does not
account for who else was on the ice and other confounding factors.  Hence, PFP
is the partial effect version of FP.

An important feature of the standard PM statistic that differs from both
\textit{for-\%} and $\beta$ or PFP is that it -- in a limited sense -- accounts
not only for player ability but also the amount that they play.  For example,
a player with a very high PM must both perform well \textit{and} maintain this
level of performance over an extended period of time (assuming that you need
to be on the ice a long time to be on the ice for many goals).  Conversely,
similarly estimated $\beta$ values for two players might hide the fact that one
of these two logs much more ice time and is thus more valuable to the
team.\footnote{Note that, due to the role of the penalty in our regularized
estimation scheme,  players with little ice time tend to have their effect
estimated at zero; thus, the difference between $\beta$ or PFP and the PPM
statistics should be less dramatic than the difference between FP and PM
statistics. In a fully Bayesian analysis, such as the {\tt reglogit} approach
discussed in our conclusion, one would be able to separate posterior
uncertainty about $\beta$ from the issue of the number of goals for which a
player is on ice; for example, the {\tt reglogit} approach yields posterior
uncertainty over a player's PPM.}  It is therefore also important to be able to
translate our partial player effects back to same scale as PM, and we do this
in the {\em partial plus-minus} (PPM). Suppose player $j$ was on the ice for
$g_{sj}$ total goals (for or against) during season $s$; then the PPM is
defined \begin{equation}\label{eq:ppm}  \text{PPM}_{sj} =
g_{sj}\text{PFP}_{sj} - g_{sj}(1-\text{PFP}_{sj}) = g_{sj}(1 -
2\text{PFP}_{sj}).  \end{equation}   Just as PFP is is the partial effect
version of FP, PPM is the partial effect analogue to PM's marginal effect.
% The same calculations apply for  post-season PFP and PPM, based upon
% $\text{PFP}_{spj} = 1/\exp[\beta_{0j} + \beta_{sj}+ \beta_{pj}]$ and where $g_{spj}$
% is the number of goals for which player $j$ was on ice during the playoffs of
% season $j$.

Table \ref{tab:goals} provides a list of top and bottom players listed by
their PPM, along with the corresponding $\beta$ effects and their standard PM.
Since these PPMs and effects are calculated for each season, players will
occur repeatedly in the table; for example, Sidney Crosby has 4 of the top 10
best player-seasons since 2002; he has been consistently the best, or near
best, player in the league.  The number one player-season since 2002 by PPM is
Peter Forsberg in 2002-2003, with a PPM of 55.5.  This is around 25\% better
than the 2nd best PPM: Crosby's 43.5 in 2009-2010.  These tabulated effects
are all calculated based upon regular season performance alone.  However,
 for all 10,000 player seasons, using goals data, we never see
enough signal to conclude that a given player was significantly better or
worse in the post-season than in the regular season.  That is, the $\hat
\beta_{pj}$ are all zero and we have no evidence of `clutch' players who
improve their play in the post-season.  At the same time, many of the
$\beta_{sj}$ {\it are} estimated at nonzero values: there is measurable signal
indicating that player performance changes across seasons.

\begin{table}[p]
    \centering\small
    \begin{tabular}{r c c c|r r r r }
        \multicolumn{8}{l}{\bf Goal-based performance analysis} \\ \\
        Rank & Player & Season  & Team & PFP & FP & PPM & PM \\ \hline
        \rule{0pt}{4ex} 
        1&PETER FORSBERG&2002-2003&COL&0.68&0.77&55.52&85\\
        2&SIDNEY CROSBY&2009-2010&PIT&0.60&0.64&43.47&60\\
        3&DOMINIK HASEK&2005-2006&OTT&0.59&0.67&42.45&80\\
        4&SIDNEY CROSBY&2008-2009&PIT&0.60&0.61&42.26&48\\
        5&SIDNEY CROSBY&2005-2006&PIT&0.60&0.62&41.86&52\\
        6&PETER FORSBERG&2005-2006&PHI&0.68&0.77&40.67&61\\
        7&PAVEL DATSYUK&2007-2008&DET&0.60&0.72&39.49&87\\
        8&PAVEL DATSYUK&2008-2009&DET&0.60&0.67&39.49&69\\
        9&SIDNEY CROSBY&2006-2007&PIT&0.60&0.72&35.62&79\\
        10&MARK STREIT&2008-2009&NYI&0.59&0.56&35.08&24\\
        11&MATT MOULSON&2011-2012&NYI&0.60&0.61&34.92&37\\
        12&LUBOMIR VISNOVSKY&2010-2011&ANA&0.58&0.66&34.52&70\\
        13&ALEX OVECHKIN&2008-2009&WAS&0.57&0.66&34.46&80\\
        14&JOE THORNTON&2009-2010&SJS&0.60&0.65&33.91&52\\
        15&JOE THORNTON&2010-2011&SJS&0.60&0.64&33.91&48\\
        16&ONDREJ PALAT&2013-2014&TAM&0.64&0.66&32.75&37\\
        17&PAVEL DATSYUK&2006-2007&DET&0.60&0.71&32.61&70\\
        18&JOE THORNTON&2002-2003&BOS&0.60&0.64&32.17&47\\
        19&JOE THORNTON&2007-2008&SJS&0.60&0.71&32.17&69\\
        20&ANDREI MARKOV&2007-2008&MON&0.57&0.60&31.9&47\\
        21&PETER FORSBERG&2003-2004&COL&0.68&0.72&31.47&39\\
        22&JOE THORNTON&2008-2009&SJS&0.60&0.67&31.21&56\\
        23&PETER FORSBERG&2006-2007&PHI&0.68&0.68&31.12&32\\
        24&PAVEL DATSYUK&2005-2006&DET&0.60&0.74&30.85&75\\
        25&ROBERT LANG&2003-2004&WAS&0.60&0.66&30.8&50\\
        \hline             \rule{0pt}{4ex}
        10184&PATRICK LALIME&2008-2009&BUF&0.43&0.44&-15.79&-15\\
        10185&JACK JOHNSON&2007-2008&LOS&0.45&0.39&-15.82&-34\\
        10186&BRETT CLARK&2011-2012&TAM&0.44&0.35&-16.93&-47\\
        10187&NICLAS HAVELID&2008-2009&ATL&0.45&0.39&-16.97&-40\\
        10188&JACK JOHNSON&2010-2011&LOS&0.45&0.53&-17.21&9\\
        10189&JACK JOHNSON&2011-2012&LOS&0.45&0.5&-17.21&-1\\
        10190&P. J. AXELSSON&2008-2009&BOS&0.41&0.49&-17.35&-1\\
        10191&BRYAN ALLEN&2006-2007&FLA&0.45&0.45&-17.9&-17\\
        10192&JACK JOHNSON&2009-2010&LOS&0.45&0.49&-19.46&-4\\
        10193&PATRICK LALIME&2005-2006&STL&0.43&0.40&-19.77&-29\\
        10194&ALEXANDER EDLER&2013-2014&VAN&0.37&0.27&-20.49&-35\\
        10195&PATRICK LALIME&2007-2008&CHI&0.43&0.49&-22.29&-4\\
        10196&TIM THOMAS&2009-2010&BOS&0.43&0.46&-24.22&-16\\
        10197&ANDREJ MESZAROS&2006-2007&OTT&0.42&0.48&-27.32&-6\\
        10198&BRYCE SALVADOR&2008-2009&NJD&0.35&0.37&-34.4&-31\\
        10199&PATRICK LALIME&2002-2003&OTT&0.43&0.58&-37.81&47\\
        10200&PATRICK LALIME&2003-2004&OTT&0.43&0.56&-37.81&37\\
        10201&NICLAS HAVELID&2006-2007&ATL&0.34&0.44&-62.64&-22\\
        10202&NICLAS HAVELID&2005-2006&ATL&0.33&0.40&-65.94&-41\\
        10203&JAY BOUWMEESTER&2005-2006&FLA&0.33&0.42&-69.62&-32\\
    \end{tabular}
    \vskip .5cm
    \caption{\label{tab:goals} Top-25 and bottom-20 player-seasons when ranked by their regular-season PPM.  }
\end{table}

The ranking in Table (\ref{hockeymod}) differs dramatically from those in
\cite{gramacy:jensen:taddy:2013}.  This occurs because we're now
\textit{controlling} for additional non-player confounding factors (e.g.,
coaching through team-season effects) and we allow the player performance to
change over time rather than be fixed at a single `career' value.   To help
intuition on why such control and model flexibility is useful, we note that
Sidney Crosby's $\beta$ effects drop significantly (he falls out of the top 5
for any season) if you do not control for the special teams effects.  This
occurs because he spends a lot of time on the penalty kill (short-handed),
which makes it easier for him to get scored upon through no fault of his own.
As another example, many of the goalies have large PPM if you do not control
for team-season effects; since the goalie is almost always on the ice, they
act as a surrogate for aggregate team performance unless you explicitly
control for it (unfortunately for Patrick Lalime, there is still enough
variation at goal to measure the effect and PPM for some goalies).

Another change from \cite{gramacy:jensen:taddy:2013} is that we are ranking
players here by PPM rather than by $\beta$; as described above, this rewards
those with more ice-time. For comparison, Table \ref{tab:pfp} ranks players by
their PFP (which is equivalent to ranking by $\beta$); the table includes both
the goal-based metrics from this section and the shot-based metrics from our
next section. While PFP and PPM are clearly related quantities, we do see some
major differences.  For example, Tyler Toffoli (ranks 9 and 10 by goal-based
PFP) was a breakout star in 2013-2014 with the Los Angeles Kings; this was his
first full season, after playing only a portion of 2012-2013 in the NHL.  As a
rookie, his ice time was relatively limited; however he clearly has talent and
this is reflected in his $\beta$  and PFP but less in his PPM.  On the
other hand, players ranked at the bottom by PPM in Table \ref{tab:goals} are those who have a negative
$\beta$ \textit{and} get a large amount of ice-time. There are many players who have
lower PFPs than Jack Johnson's 0.45 (e.g., John McCarthy at 0.38 and Thomas Pock
at 0.40), but they do not get to play as much and thus don't show up in our
bottom 20.

\begin{table}[t]
    \centering\small
    \begin{tabular}{r | c c c r  | c c c r  }
        \multicolumn{9}{l}{\bf PFP player rankings}\\
        \multicolumn{5}{c}{goal-based} & \multicolumn{4}{|c}{Corsi-based}\\
        Rank & Player & Season & Team & PFP  & Player & Season  & Team & PFP  \\ 
        \hline\rule{0pt}{4ex} 
        1&PETER FORSBERG&2002-2003&COL&0.68&DAVID VAN DER GULIK&2010-2011&COL&0.64\\
        2&PETER FORSBERG&2005-2006&PHI&0.68&DAVID BOOTH&2012-2013&VAN&0.63\\
        3&PETER FORSBERG&2003-2004&COL&0.68&DANIEL SEDIN&2012-2013&VAN&0.62\\
        4&PETER FORSBERG&2006-2007&PHI&0.68&ALEXANDER SEMIN&2003-2004&WAS&0.61\\
        5&PETER FORSBERG&2007-2008&COL&0.68&DANIEL SEDIN&2010-2011&VAN&0.60\\
        6&PETER FORSBERG&2010-2011&COL&0.68&MIKHAIL GRABOVSKI&2010-2011&TOR&0.60\\
        7&ONDREJ PALAT&2013-2014&TAM&0.64&DANIEL SEDIN&2007-2008&VAN&0.60\\
        8&ONDREJ PALAT&2012-2013&TAM&0.64&DANIEL SEDIN&2008-2009&VAN&0.60\\
        9&TYLER TOFFOLI&2013-2014&LOS&0.63&DANIEL SEDIN&2011-2012&VAN&0.60\\
        10&TYLER TOFFOLI&2012-2013&LOS&0.63&PATRIK ELIAS&2010-2011&NJD&0.60\\
        11&VINCENT LECAVALIER&2006-2007&TAM&0.61&SIDNEY CROSBY&2013-2014&PIT&0.60\\
        12&VINCENT LECAVALIER&2003-2004&TAM&0.61&DANIEL SEDIN&2009-2010&VAN&0.60\\
        13&SIDNEY CROSBY&2009-2010&PIT&0.60&JUSTIN WILLIAMS&2010-2011&LOS&0.60\\
        14&SIDNEY CROSBY&2008-2009&PIT&0.60&DANIEL SEDIN&2013-2014&VAN&0.60\\
        15&SIDNEY CROSBY&2005-2006&PIT&0.60&PATRIC HORNQVIST&2013-2014&NSH&0.60\\
        16&PAVEL DATSYUK&2007-2008&DET&0.60&PAVEL DATSYUK&2012-2013&DET&0.60\\
        17&PAVEL DATSYUK&2008-2009&DET&0.60&ALEX STEEN&2011-2012&STL&0.60\\
        18&SIDNEY CROSBY&2006-2007&PIT&0.60&BRAD RICHARDSON&2011-2012&LOS&0.60\\
        19&MATT MOULSON&2011-2012&NYI&0.60&ERIC FEHR&2008-2009&WAS&0.60\\
        20&JOE THORNTON&2009-2010&SJS&0.60&TYLER TOFFOLI&2013-2014&LOS&0.60\\
    \end{tabular}
    \vskip .25cm
    \caption{\label{tab:pfp} Top 20 player-seasons by goal and Corsi-based PFP.}
\end{table}

\section{Analysis: comparison to shot-based metrics}
\label{sec:shots}

The analysis above is built around the event of a `goal'; this is the most
reasonable baseline analysis, as it removes any subjectivity about whether or
not the statistics are related to team performance -- you score more you win.
However, it has recently become popular in hockey analysis to consider
alternative metrics that are built from shots and other events; see
\cite{vol10} for a review. The most popular of such statistics is Corsi, which
counts the number of events that are goals, shots on goal, missed shots, or
blocked shots.  Fenwick is another statistic; it is Corsi but without counting
blocked shots.  Although we have seen no evidence that Corsi or Fenwick events
are more useful in predicting team performance than goal-based metrics, they
do offer a big advantage to the statistician: they lead to a larger sample
size, so that you can hopefully better identify the competing influences of
different players and confounding factors. Our data contain
$n_{c}=1,329,679$ Corsi events and $n_{f}=1,034,154$ Fenwick events; this is
an order of magnitude more events than the $n_g=69449$ goals.

\begin{table}[p]
    \centering\small
    \begin{tabular}{r c c c| r r r r  }
        \multicolumn{8}{l}{\bf Corsi-based performance analysis}\\ \\
        Rank & Player & Season & Team & PFP & FP & PPM & PM   \\ 
        \hline\rule{0pt}{4ex} 
        1&DANIEL SEDIN&2010-2011&VAN&0.60&0.65&615.14&876\\
        2&ERIC STAAL&2008-2009&CAR&0.58&0.59&605.41&619\\
        3&MIKHAIL GRABOVSKI&2010-2011&TOR&0.60&0.57&597.05&465\\
        4&JOE THORNTON&2011-2012&SJS&0.59&0.61&596.37&742\\
        5&ALEX OVECHKIN&2009-2010&WAS&0.59&0.66&575.72&1047\\
        6&DANIEL SEDIN&2007-2008&VAN&0.60&0.63&562.11&685\\
        7&DANIEL SEDIN&2008-2009&VAN&0.60&0.62&547.83&680\\
        8&RYAN KESLER&2010-2011&VAN&0.58&0.59&530.05&649\\
        9&SIDNEY CROSBY&2009-2010&PIT&0.57&0.62&517.86&815\\
        10&DANIEL SEDIN&2011-2012&VAN&0.60&0.67&510.16&880\\
        11&HENRIK ZETTERBERG&2011-2012&DET&0.58&0.60&497.04&596\\
        12&CLAUDE GIROUX&2010-2011&PHI&0.58&0.56&487.53&347\\
        13&ZACH PARISE&2008-2009&NJD&0.58&0.64&486.45&843\\
        14&JOE THORNTON&2010-2011&SJS&0.58&0.60&482.72&647\\
        15&ALEX STEEN&2010-2011&STL&0.59&0.61&475.5&561\\
        16&LUBOMIR VISNOVSKY&2010-2011&ANA&0.56&0.56&474.91&446\\
        17&ERIC STAAL&2010-2011&CAR&0.56&0.56&473.92&415\\
        18&JUSTIN WILLIAMS&2011-2012&LOS&0.59&0.63&471.53&717\\
        19&ALEX OVECHKIN&2007-2008&WAS&0.56&0.65&463.14&1094\\
        20&PATRIK ELIAS&2010-2011&NJD&0.60&0.60&461.75&461\\
        21&SIDNEY CROSBY&2013-2014&PIT&0.60&0.61&459.92&480\\
        22&DUSTIN BYFUGLIEN&2010-2011&ATL&0.56&0.60&456.04&705\\
        23&JAROMIR JAGR&2007-2008&NYR&0.58&0.65&455.78&911\\
        24&ALEX OVECHKIN&2008-2009&WAS&0.56&0.64&455.46&1065\\
        25&JASON BLAKE&2008-2009&TOR&0.58&0.55&454.7&278\\
        \hline\rule{0pt}{4ex}
        10605&MIKE COMMODORE&2008-2009&CBS&0.43&0.42&-447.91&-537\\
        10606&SCOTT HANNAN&2011-2012&CGY&0.42&0.40&-451.04&-591\\
        10607&CHRIS PHILLIPS&2007-2008&OTT&0.43&0.40&-454.09&-644\\
        10608&JAY BOUWMEESTER&2005-2006&FLA&0.44&0.46&-457.01&-305\\
        10609&KARLIS SKRASTINS&2008-2009&COL&0.43&0.38&-457.54&-754\\
        10610&KARLIS SKRASTINS&2009-2010&DAL&0.42&0.39&-464.49&-655\\
        10611&MATTIAS OHLUND&2008-2009&VAN&0.42&0.47&-465.36&-212\\
        10612&MATTIAS OHLUND&2006-2007&VAN&0.43&0.48&-470.03&-147\\
        10613&SCOTT HANNAN&2008-2009&COL&0.43&0.38&-478.83&-788\\
        10614&DOUGLAS MURRAY&2009-2010&SJS&0.42&0.47&-486.16&-184\\
        10615&SCOTT HANNAN&2007-2008&COL&0.42&0.42&-507.7&-504\\
        10616&FILIP KUBA&2011-2012&OTT&0.42&0.49&-509.74&-77\\
        10617&NICLAS HAVELID&2007-2008&ATL&0.41&0.35&-516.86&-883\\
        10618&JOHNNY ODUYA&2008-2009&NJD&0.42&0.51&-522.4&51\\
        10619&DOUGLAS MURRAY&2010-2011&SJS&0.40&0.48&-540.83&-117\\
        10620&DION PHANEUF&2006-2007&CGY&0.42&0.49&-552.69&-42\\
        10621&NICLAS HAVELID&2008-2009&ATL&0.40&0.40&-562.65&-604\\
        10622&SERGEI GONCHAR&2006-2007&PIT&0.42&0.52&-586.55&174\\
        10623&PAUL MARTIN&2008-2009&NJD&0.39&0.55&-695.83&283\\
        10624&BRYCE SALVADOR&2008-2009&NJD&0.32&0.42&-912.17&-407\\
    \end{tabular}
    \vskip .5cm
    \caption{\label{tab:corsi} Top 25 and bottom 20 players by Corsi-based PPM.}
\end{table}

The standard way to report Corsi and Fenwick for a given player is as the
\textit{for-\%} (FP) described above.  Again, since the FP score does not
reward players for the amount of time that they spend on the ice, we also
consider both Corsi and Fenwick versions of the plus-minus statistic. Of
course, all of these statistics -- Corsi-FP, Corsi-PM, etc.~-- measure
marginal effects.  They are thus subject to the same criticisms as the
original PM: they fail to control for the influence of other players and
confounding factors, and are thus less useful than a partial effect for
predicting and measuring player performance.  However, we can apply the exact
same regression analysis that we've used above for goal events to derive
\textit{partial} versions of the Corsi and Fenwick statistics: simply replace
$y_i$ with a response calculated from Corsi or Fenwick events.  For example, a
Corsi regression applies the model as in (\ref{hockeymod}) but for response
$y_i=+1$ if the event was a Corsi event (shot, goal, blocked shot) by the home
team and $y_i=-1$ if it was a Corsi event by the away team. The partial \textit{for-\%}
and plus-minus formulas of (\ref{eq:pfp}-\ref{eq:ppm}) can similarly be applied to 
obtain Corsi-PPF and Corsi-PPM values.

The results for regular season Corsi-based performance analysis are in Table \ref{tab:corsi}
and on the right side of Table \ref{tab:pfp}.  Comparison to the goal-based
rankings shows a distinctly different set of players are at both the top and
bottom.  For example, Daniel Sedin is a prominent player who ranks highly in
multiple seasons under Corsi-PPM but does not appear in the top-20 for
goals-PPM (his best goals-PPM is a still respectable 19.45 in 2010-2011, which
ranks 152$^{nd}$ across all player-seasons).  At this point we are not looking
to argue for either the goal or Corsi based metrics as `best'; however, the
fact that they do differ dramatically should be a bit troubling for those who
wish to focus exclusively on Corsi statistics (since only goal differentials
dictate who wins the game).  In the next section, we consider a comparison
between all of our metrics and an outside measure of player quality: salary.

\section[Salary and Performance]{Analysis: the relationship between salary and performance}
\label{sec:salary}

\begin{figure}[t]
    \centering
    \begin{subfigure}[t]{0.5\textwidth}
        \centering
        \includegraphics[height=2.25in]{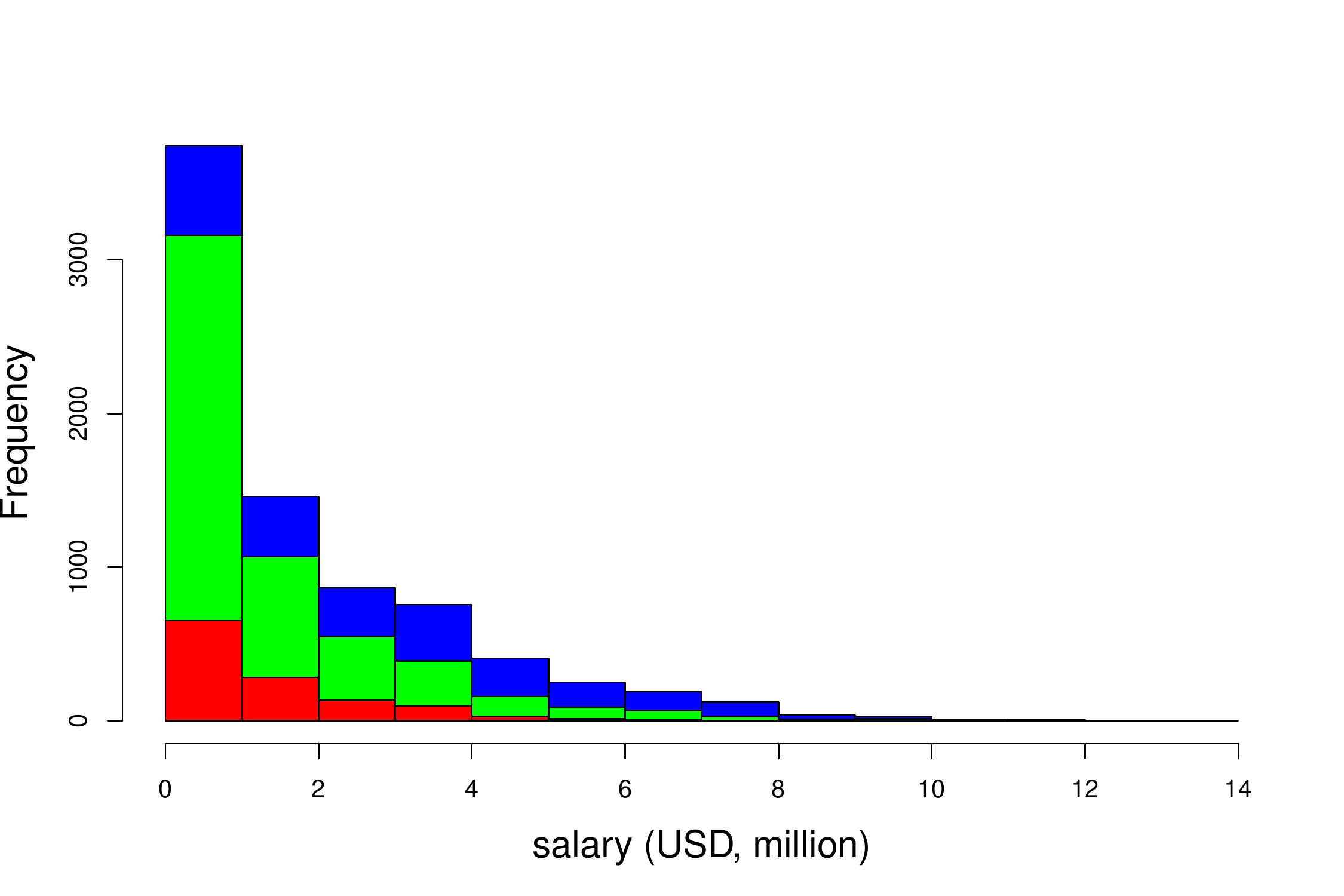}
        \caption{Goals-based}
    \end{subfigure}%
    ~ 
    \begin{subfigure}[t]{0.5\textwidth}
        \centering
        \includegraphics[height=2.25in]{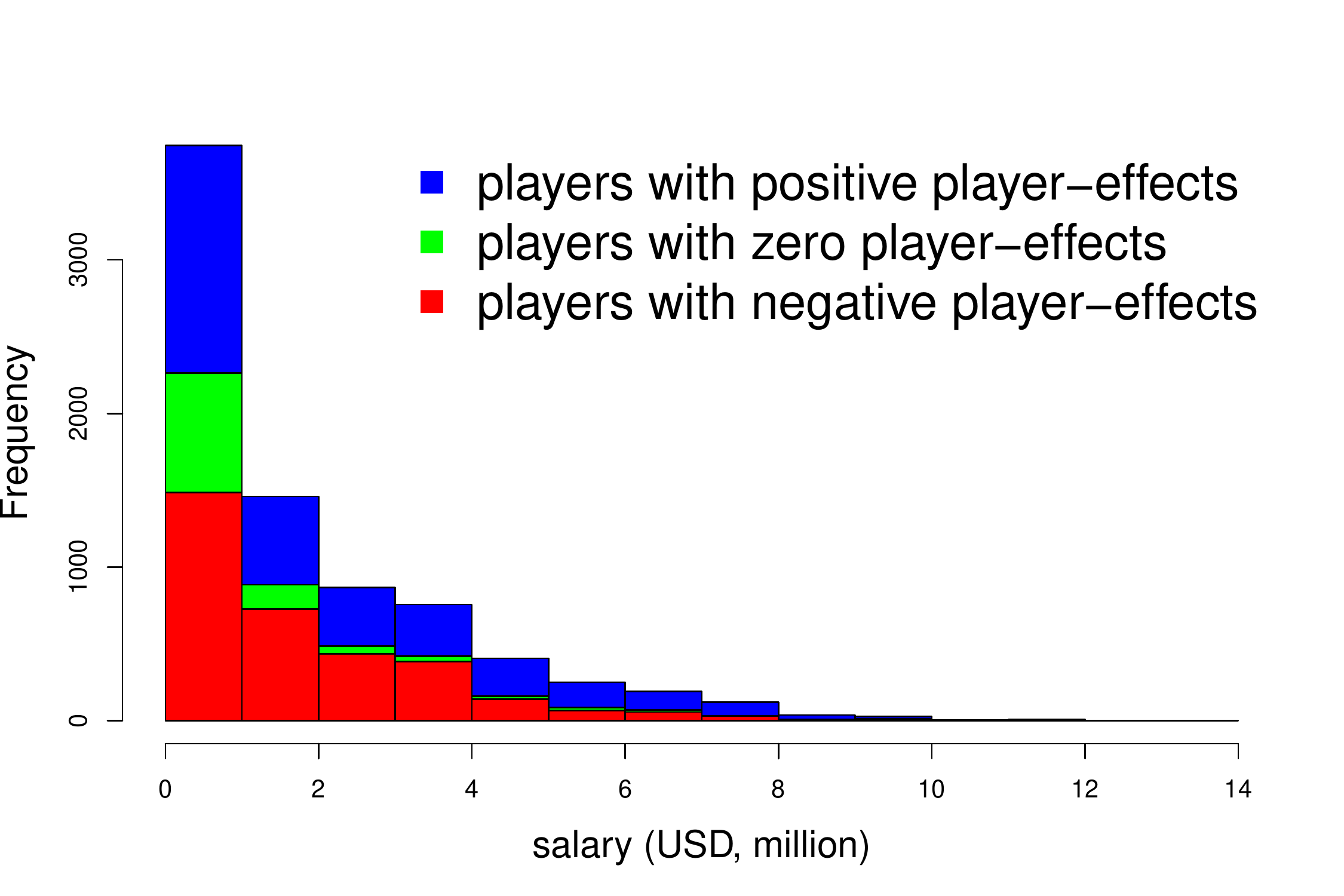}
        \caption{Corsi-based}
    \end{subfigure}
    \caption{ \label{fig:salhist} Distribution of estimated player effects and salaries over all 11 seasons.}
\end{figure}

In our final analysis section, we consider how our partial effect statistics
relate to the \textit{market} value of a player's worth as represented by
their annual salary.  The salary numbers are obtained through a combination of
the databases maintained at {\sf blackhawkzone.com} and {\sf
hockeyzoneplus.com}; we are able to obtain annual salaries for 80\%
of the player-seasons in our dataset, including almost all player-seasons
where the player was on ice for more than a couple of goals. The histograms in
Figure \ref{fig:salhist} show, for the salary distribution over all 11
seasons, how the estimated player effects are distributed between negative,
neutral (zeros), and positive.  Results are shown for both Goal and Corsi
based regressions.  In both cases, the ratio of positive to negative effects
increases with salary.  The main difference between the two plots is that
fewer players have zero estimated effects under Corsi -- this is a result of
its much larger event sample.

\begin{figure}[tbh]
    \centering
    \includegraphics[width=\textwidth]{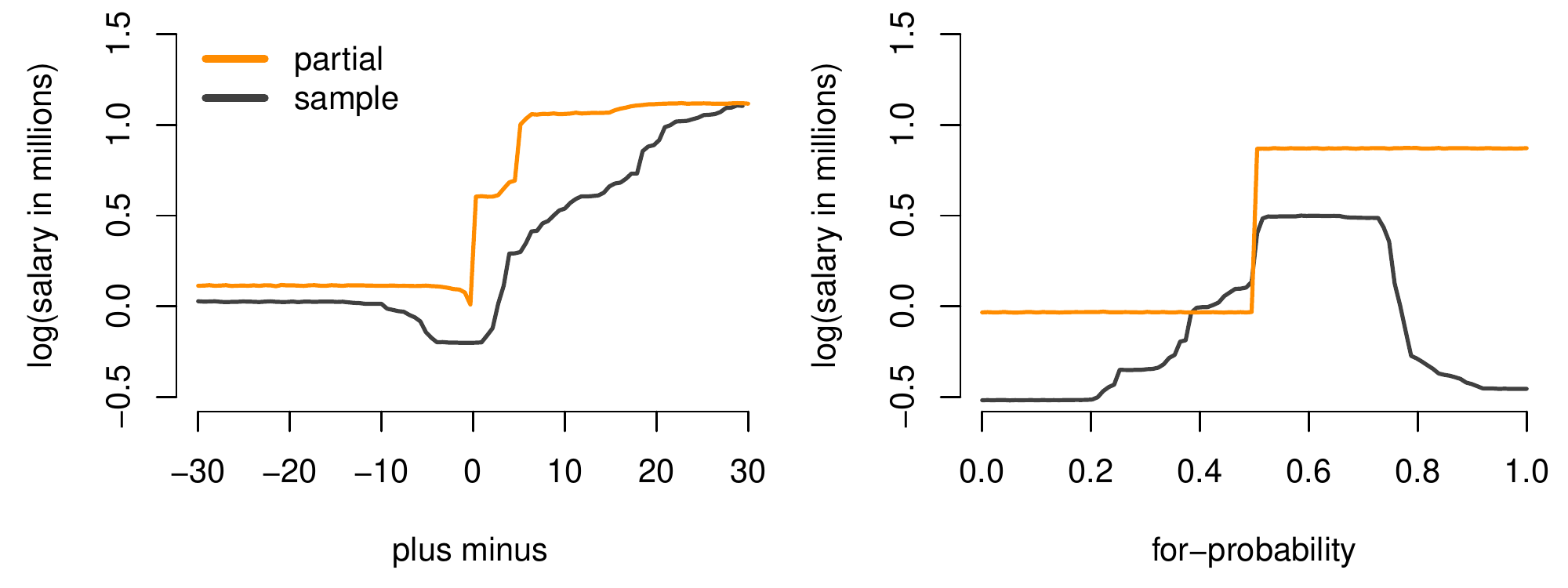}
    \caption{Nonparametric regression for log average salary (per player) onto their average performance metric: PPM, PM, PFP, and FP.  }\label{fig:salaryreg}
\end{figure}

Figure \ref{fig:salaryreg} takes a deeper look at the relationship between
salary and performance for two of our goal-based metrics: the goal
plus-minuses (PM and PPM) and for-percentages (FP and PFP). For each metric,
we use nonparametric Bayesian regression to fit expected log salary as a
function of that metric.   In particular, we apply the {\tt tgp} package
\cite{gramacy_tgp:_2007,gramacy_categorical_2010} to obtain posterior means
for the Bayesian regression trees of
\cite{chipman_bayesian_2002}.  The trees are fit to salary and performance
data that has first been aggregated by player, such that these surfaces
represent the expected log average salary (per player) conditional upon their
average performance across the years that they played.  This aggregation is
done  to minimize dependence between observations, and because we assume that
a player's salary is not determined by a single season.

The surfaces in \ref{fig:salaryreg} expose some interesting differences
between the partial and marginal performance statistics.  In the plus-minus
case (PM and PPM), the two curves are similar -- they both show little
relationship between salary and performance for negative plus-minus, while
salaries rise with positive performance.   However, there is a larger jump up
from zero for PPM  than for PM.  This occurs because the regression estimation
only assigns a positive non-zero effect for statistically significant
performances, so that players with small but positive PM values have their PPM
shrunk to zero.  The difference between FP and PFP is more dramatic.  In the
case of marginal FP, the salaries are highest for players with with FP in the
middle of the range (near and above 0.5).  This occurs because FPs outside of
that range occur only for players with little ice-time.  In contrast, the PFP
relationship with salary is very simple: salaries are low for PFP below 0.5,
and high above that number.  If you make it more likely than not that a
goal-scored is a goal-for, you can expect to make more money.

Finally, we close with a look at the `highest value' players in the 2013-2014
season: those for whom their salary was low relative to their goal-based PPM.
Table \ref{tab:valueplayers} lists the top 20 players by goal-PPM/salary --
that is, the top players measured by their goals-added-per-dollar.  The
cheapest player contribution, by a massive margin, is from Ondrej Palat of
Tampa Bay.  Palat was an inexpensive 7th round draft pick in
2011, making \$500 thousand per year. After spending two seasons in the
minors, he moved up to the NHL in 2013-2014 and had a season good enough to be
nominated for rookie-of-the-year.  The Lightning re-signed Palat to a new
contract in 2014; he now averages around \$3.5 million per year.

\begin{table}[htb]
    \centering\small
    \begin{tabular}{r c c r   }
        Rank & Player & Team & Goals per million  \\ 
        \hline\rule{0pt}{4ex} 
        1&ONDREJ PALAT&TAM&58.27\\
        2&RYAN NUGENT-HOPKINS&EDM&19.81\\
        3&GABRIEL LANDESKOG&COL&16.74\\
        4&TYLER TOFFOLI&LOS&16.72\\
        5&GUSTAV NYQUIST&DET&9.08\\
        6&JADEN SCHWARTZ&STL&8.43\\
        7&ERIC FEHR&WAS&7.51\\
        8&ANDREW MACDONALD&NYI&7.48\\
        9&BENOIT POULIOT&NYR&6.43\\
        10&BRAD BOYES&FLA&6.01\\
        11&TOMAS TATAR&DET&5.83\\
        12&AL MONTOYA&WPG&5.79\\
        13&BRANDON SAAD&CHI&5.5\\
        14&FRANS NIELSEN&NYI&5.5\\
        15&JAROMIR JAGR&NJD&4.73\\
        16&LOGAN COUTURE&SJS&4.7\\
        17&RADIM VRBATA&PHO&4.4\\
        18&DAVID PERRON&EDM&4.1\\
        19&HENRIK LUNDQVIST&NYR&3.76\\
        20&ANDREI MARKOV&MON&3.5\\
    \end{tabular}
    \vskip .5cm
    \caption{\label{tab:valueplayers} Top 20 value players as ranked by PPM/salary. }
\end{table}

\section{Conclusion}\label{sec:conclusion}

We have provided a sketch for how modern techniques in regularized logistic
regression, developed originally to address challenging large-scale problems
in genetics, finance, and text mining, can be used to calculate partial player
effects in hockey.  We have argued that such  partial effects are a
better measure of player ability compared to the classic plus-minus statistic,
and have the benefit of being interpretable on the same scale as plus-minus.
 We have shown
how the framework is flexible, allowing one to control for many aspects of
situational play (special teams, overtime, playoffs), and personnel/season
(coaches, salaries, season-years).  A comparison was provided to the popular
Corsi and Fenwick alternatives to plus-minus, and we argued that a recent
emphasis in the literature on shots (and blocked shots), does not in general
compare favorably to the traditional goals focus in this framework.

Our development has focused on point-estimation via the {\tt gamlr} package,
which infers parameters under $L_1$ penalization.  Another software package
offering similar features is {\tt glmnet}
\cite{zou_regularization_2005}.  The difference between {\tt gamlr} and {\tt
glmnet} is in what options they provide on top of the standard $L_1$ penalty.
As detailed in \cite{taddy_one-step_2015}, including an example analysis of
this same hockey data, {\tt gamlr} provides a `gamma-lasso' algorithm for
\textit{diminishing bias} penalization: the penalty on coefficients
automatically diminishes for strong signals.  If you believe that the current
analysis has over-shrunk the influence of, say, total stars like Sidney Crosby
or Pavel Datsyuk, then {\tt gamlr} and \cite{taddy_one-step_2015} will offer a
preferable analysis framework.  On the other hand, if you think that all
players should be shrunk \textit{closer} to zero -- perhaps you believe that
the current results over-state the effect of a few stars -- then the elastic
net penalization scheme of \cite{zou_regularization_2005} and {\tt glmnet}
will be preferable.  In either case, simple $L_1$ penalization provides a
useful reference baseline analysis.

Beyond changes to the penalty specification, we think that there can be
considerable value in moving from point-estimation to a fully Bayesian
analysis. \cite{gramacy:jensen:taddy:2013} included exploration of the player
effect posterior in their earlier analysis of a related model.  They apply the
{\tt reglogit} package \cite{reglogit} for {\sf R}, which implements the Gibbs
sampling strategy from \cite{gra:pols:2012}. The software takes advantage of
sparse matrix libraries ({\tt slam} \cite{slam}), and is multi-threaded via
{\tt OpenMP} to engage multiple processors simultaneously. It combines two
scale-mixture of normals data-augmentation schemes, one for the logit
\cite{holmes:held:2006} and one for the Laplace prior \cite{park:casella:2008}.
Obtaining $T$ samples from the full posterior, is straightforward using the
following {\sf R} code
\begin{verbatim}
bfit <- reglogit(T=T, y=Y, X=X, normalize=FALSE)
\end{verbatim}
The full posterior sample for $\beta$, residing in {\tt bfit}, is available
for calculation of posterior means and covariances of player effects and other
posterior functionals relevant to player performance. For example,
\cite{gramacy:jensen:taddy:2013} use the posterior \textit{probability} that
one player is better than another as a basis for ranking players, and can even
provide posterior credible intervals around these rankings.\footnote{see.,
e.g,
\url{https://github.com/Tad dyLab/hockey/blob/master/results/blog/logistic_pranks_betas.csv}
}  This information can be used to construct teams of players under budget constraints and
subsequently describe the probability that those teams will score more goals
than their opponents.  

Another option is to depart from the restrictions of linear modeling.  Anecdotally, some in the
sports analytics community (not just in hockey) have embraced a framework
built around random forests
\cite{breiman:2001}.  An advantage of decision trees, on which random forests
are based, is that they naturally explore interactions between predictors --
e.g., between players and other effects in the hockey analysis.  The bagging
procedure -- averaging across many trees -- provides a mechanism for avoiding
over-fit: structure that is not persistent across trees is eliminated by the
averaging. Such work also fits with our above advocacy of fully Bayesian
analysis:
 \cite{taddy:eta:2015} describes random forests as approximating a Bayesian
 nonparametric posterior over trees, while  Bayesian additive regression trees
 (BART
\cite{ChipGeorMcCu2010}) provide an alternative tree-based scheme that can be
extended to logistic regression via the  latent variable techniques in
\cite{gra:pols:2012}.   Finally, the proportional hazards model
\cite{ThoVenJen12}, mentioned above in Section \ref{sec:regression},  attempts to reproduce more completely the stochastic
processes behind scoring in a hockey game. Their fully Bayesian analysis
accounts for a wide set of game information, including the time-on-ice
information that we are only roughly accounting for in our PPM statistic.

Regardless of these and other possible complex extensions, we argue strongly
that our simple $L_1$ penalized logistic regression has much to recommend it.
The model is very simple to interpret and relies upon minimal restrictive
assumptions on the process of a hockey game.    Our measures are also much
faster to compute than any of the alternatives. These qualities make
sophisticated real-time analysis of player effects possible as games and
seasons progress.

\bibliographystyle{plain}
\bibliography{hockey}
\end{doublespace}

\end{document}